\documentstyle[times,pramana,epsf,floats]{ias}
\begin{document}
\mark{{Leptogenesis}{E.A. Paschos}}
\title{Leptogenesis}

\author{E.A. Paschos}
\address{University of Dortmund, Institute of Physics,
44221 Dortmund, Germany}
\abstract{I present the theoretical basis for 
Leptogenesis and its implications for the structure of
the universe. It is suggested that density fluctuations
grow during the transition period and remnants of this
effect should be sought in the universe.  The relation
between theories with Majorana neutrinos and low energy
phenomena, including oscillations, advanced considerably
during the past two years with a consistent picture
developed in several models.}

\maketitle
\section{Introduction}
The discovery of neutrino oscillations signals the
extension of the standard model with the introduction
of right--handed neutrinos.  The evidence so far shows
the mixing of $\nu_e\to\nu_\mu$ (solar) and
$\nu_{\mu}\to\nu_{\tau}$ (atmospheric) with very
small mass differences and large mixing angles.
The mixing of $\nu_e\to\nu_{\tau}$ has not been observed
yet and consequently the mixing angle must be much
smaller.
The oscillations require right--handed components
for the states in order to produce neutrino masses.
Introducing the right--handed components of neutrinos
leads to several new issues to be investigated.
\begin{itemize}
\item[1)] Are the neutrinos Dirac or Majorana
particles?  In other words, should we introduce
only Dirac type terms in the mass matrices
or also Majorana mass terms. When we follow the
philosophy that all allowed mass terms must be
included, then Majorana terms must be present.
\item[2)] Is the mixing observed so far related
to other phenomena?  An attractive possibility
is related to the generation of a lepton asymmetry
in the early universe, which was subsequently
converted into a Baryon asymmetry.
\item[3)] Is it possible to find evidence for
the Majorana nature of neutrinos?
\item[4)] Are the low energy phenomena related
to the high energy theory we need for Leptogenesis?
\end{itemize}
This field has become very interesting because
there are many unanswered questions, some of 
which can be investigated experimentally.  I will
address some of these issues in this talk.

\section{Majorana Neutrinos}
The neutrinos participate in their observed
interactions as 
left--handed particles.  The right--handed component
of the Dirac spinor was intentionally left out believing that 
neutrinos were massless.  The mixing phenomena 
require the introduction of right--handed components.
With them we have the choice of two mass terms
\begin{eqnarray}
\bar{\nu}_L N_R\quad & {\rm Dirac} \quad &
  \Delta L=0\\
\bar{N}_R^C N_R\quad & {\rm Majorana} \quad &
  \Delta L=2 \,.
\end{eqnarray}
One may think of introducing also the term
\begin{equation}
\quad\quad\quad\quad\quad\quad
\bar{\nu}_L^C\nu_L
\end{equation}
but this carries weak isospin of two units and must
couple to a Higgs triplet which is absent in the
standard model. It is natural to introduce the terms
in (1) and (2) and look for solutions of the mass
matrix.  In general, we obtain solutions
\begin{equation}
\quad\quad\quad\quad\quad\quad
\psi=\frac{1}{\sqrt{2}} (N_R+N_R^C)
\end{equation}
which are Majorana states and even under charge
conjugation (C). Introducing bare Majorana mass
terms, like the one in eq.~(2), one obtains
physical states which are C--eigenstates.  In 
theories with Majorana mass terms it is possible 
to introduce interactions with scalar particles,
like
\begin{equation}
\quad\quad\quad\quad\quad\quad
 h_{ij}\bar{\ell}_Lî\phi\, \psi j
\end{equation}
where $\ell_L^i$ is the left--handed doublet (leptons),
$\phi$ the ordinary Higgs doublet and $h_{ij}$ the
coupling constant with $i$ and $j$ referring to 
various generations.
The result is that such interactions together with
the mass terms produce physical states which are neither
C-- nor CP--eigenstates.

This property was emphasized in the article by Fukugida
and Yanagida \cite{ref1} that the decays of heavy
Majorana states generate a lepton asymmetry.  Later on,
it was observed by Flanz, Sarkar and myself \cite{ref2}
that the construction of the physical states contains
an additional lepton asymmetry.  
In the latter case, the situation is analogous to the 
$K^0$ and $\bar{K}^0$ states, which mix through the box 
diagrams and produce $K_L$ and $K_S$ as physical states. 
The mixing of the neutrinos originates from fermionic
self--energies which contribute to the mass matrix of
the particles.  The final result is the creation of
an asymmetry in the decays of heavy Majorana particles,
which depends on a modified Dirac term $\tilde{m}_D$ of 
the mass matrix (the modification is discussed in 
section 5).  The final result is
\begin{eqnarray}
\varepsilon & = & \frac{\Gamma(N_{R_i}\to\ell)-
   \Gamma(N_{R_i}\to\ell^c)}
   {\Gamma(N_{R_i}\to\ell)+
    \Gamma(N_{R_i}\to\ell^c)}\\
&  = & \frac{1}{8\pi v^2}
     \frac{1}{(\tilde{m}_D^+ \,\tilde{m}_D)_{11}}
\sum_{j=2,3} {\mathrm{Im}}
 \left( (\tilde{m}_D^+\,\tilde{m}_D)_{1j}^2\right)f(x)\nonumber
\end{eqnarray} 
with
$f(x) = \sqrt{x}\left\{ \frac{1}{1-x}+1-(1+x)\ln
    \left(\frac{1+x}{x}\right)\right\}$,
$x=\left(\frac{M_j}{M_1}\right)^2$, $M_1$ the mass
of the lightest Majorana neutrino and $v$ the vacuum
expectation value of the standard Higgs.  The term
$\left(\frac{1}{1-x}\right)$ comes from the mixing of
the states \cite{ref2} and the rest from the interference of 
vertex corrections with Born diagrams \cite{ref1}. 
The above formula is an approximation for the case 
when the two masses are far apart.  In case they are
close together there is an exact formula, showing 
clearly a resonance phenomenon \cite{ref3} from 
the overlap of the two wave functions.
The origin of the asymmetry has also been studied in
field theories \cite{ref4} and supersymmetric models 
\cite{ref5,ref6}. The purpose of these articles,
especially \cite{ref4}, was to justify the formalism
and eliminate some objections.

\section{(B--L)--Asymmetry}
According to the scenario described above, the 
(B--L) quantum number is violated easily through
Majorana neutrinos.  We assign to particles lepton
 and baryon quantum numbers, as follows:
$n_B=1/3$ for each baryon, $n_L=+1$ for each 
lepton and the negative numbers for their antiparticles.
Then there is a combined number
\begin{displaymath}
n_C=3n_B-n_L=B-L
\end{displaymath}
which is conserved.  More explicitly we assign
\begin{eqnarray}
n_L=-1 & \quad\quad & n_C=+1 \quad 
{\rm for}\,{\rm antimuons}\quad 
\mu^+\quad{\rm and}\quad\bar{\nu}_{\mu}\nonumber\\
n_L=+1 & \quad\quad & n_C=-1 \quad
{\rm for}\,{\rm muons}\quad\quad
\mu^-\quad{\rm and} \quad {\nu}^{\mu}\nonumber\\
n_B=+1 & \quad\quad & n_C=+1 \quad %\quad\!\!
{\rm for}\,{\rm baryons:}\,\,{\rm protons}\,{\rm and}
\,{\rm neutrons}\nonumber\\
n_B=-1 & \quad\quad & n_C=-1
 \quad
{\rm for}\,{\rm antiprotons}\,{\rm and}\,
{\rm antineutrons}
\end{eqnarray} 
Under laboratory conditions processes involving 
violation of $n_B$ and $n_L$ play a negligible 
role, but they were important during the early
state of the universe.  Violation of $n_B$ will 
produce proton decays which so far have been shown 
to be
small.  It was thought that Grand Unified Theories
(GUT) could violate the (B--L) or the B quantum
number.  However, there is a theorem which says
that the decays of heavy Gauge Bosons into quarks
and leptons with the conventional quantum numbers
involve operators of dimensions higher than six
\cite{ref7}.  
This suppresses proton decay and the generation
of a baryon asymmetry in GUT.  The standard model
obeys this rule, but it has another property:
topological solutions of the theory (sphalerons)
conserve to a high degree of accuracy B--L but
violate B+L \cite{ref8}.  Thus there is the attractive 
possibility to generate a net (B--L) in the decays
of heavy Majorana particles and subsequently convert
a fraction of it into a baryon asymmetry.  This
scenario was named Leptogenesis and presents an
attractive possibility, perhaps the only viable
one, for Baryon generation.

\section{Lepton Asymmetric Universe}
As mentioned already, a lepton asymmetry is generated
either in the decays or the construction of the
physical states.  The two possibilities may be
distinguished by their consequences. The C-- and
CP--asymmetric states will be physical and propagating
states, if during their life times they interact many
times with particles in their surroundings.
A typical interaction is shown in Fig.\ 1 with the 
two couplings being present in the theory.

%Figure 1
\begin{figure}[htbp]
\epsfxsize=4cm
\centerline{\epsfbox{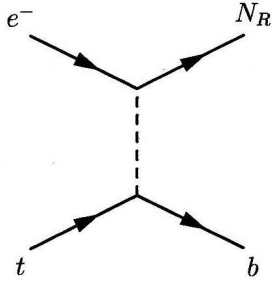}}
\caption{A typical scattering.}
\label{fig:figure1}
\end{figure}

Taking the density of states in the early universe to be
$n=\frac{2}{\pi^2} T^3$
and calculating the cross section at an energy $E=T$, 
we obtain
\begin{equation}
n\cdot \sigma\cdot u = \frac{|h_t|^2|h_{\ell}|^2}
                             {8\pi^3}\, T
\end{equation}
with $\sigma$ the cross section, $u$ their relative
velocity of the particles and $h_t$, $h_{\ell}$ the 
couplings of the
Higgses to quarks and leptons, respectively.
At that early time the decay width of the moving leptons 
with mass $M_N$ is
\begin{equation}
\Gamma_N = \frac{|h_{\ell}|^2}{16\pi}\frac{M_N^2}{T}\, ,
\end{equation}
consequently
\begin{equation}
\frac{n\cdot\sigma\cdot u}{\Gamma_N}=\frac{2}{\pi^2}\, 
|h_t|^2\left(\frac{T}{M_N}\right)^2\, .
\end{equation}

Thus, at an early stage of the universe with $T\gg M_N$,
when the mixed states are created, they live long enough 
so that
in one life--time they have many interactions with
the surroundings.  They are incoherent states.
As the universe started deviating from 
thermal equilibrium, it became lepton asymmetric and
the transition period lasted for a
relatively long time \cite{ref9,ref10}.  Estimates for the decays of the
lightest Majorana \cite{ref9} and the development of the lepton asymmetry 
are shown in figure 2.

The total asymmetry is given by
\begin{equation}
Y = D \epsilon
\end{equation}
with $D$ a dilution factor and $\epsilon$ given in eq.~(6). The dilution
factor is obtained from solutions of the Boltzmann equations \cite{ref9},
 which depend on $\kappa = \frac{\Gamma}{H}$ with $\Gamma$ the decay width and
 $H$ the Hubble constant. The asymmetry starts from zero and grows reaching 
eventually a constant value asymptotically.

%Figure 2
\begin{figure}[htbp]
\epsfxsize=8cm
\centerline{\epsfbox{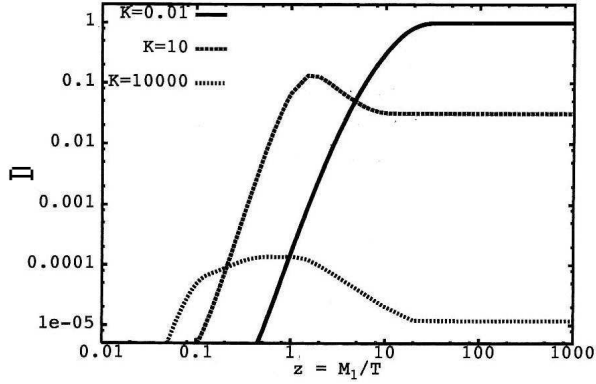}}
\caption{Dilution factor $D$.}
\label{fig:figure2}
\end{figure}
%\vspace{5.5cm}
%\begin{center}
%Figure 2.  Dilution factor $D$
%\end{center}

The figure indicates that the heavy states are
incoherent from a temperature $T_h=10\, M_1$
down to a temperature $T_l=M_1/10$.  
During this period the right--handed neutrinos are
non--relativistic and will cause fluctuations to
grow.  Heavy neutrinos, $M_N\sim 10^8$ GeV, have an
interaction length and an interaction horizon. 
Fluctuations within the interaction horizon will be washed out, 
but fluctuations at larger scales will grow.  Matter 
within the 
larger region will continue gathering together, 
forming gravitational wells which attract more 
matter. 

As the temperature of the universe approached the
mass of the lightest right--handed neutrinos the
recombination of lighter particles to Majorana
states ceased.  From the decays of the Majorana
states comes the generation of a lepton asymmetry.
An excess of
leptons survives all the way down to the electroweak 
phase transition.  During their journey from the 
decay of the Majoranas to the electroweak phase 
transition a part of the leptons was transformed
into a baryon excess.  With the conversion of
leptons into quarks, density fluctuations which
were formed during the transition period and later
on, are transformed into fluctuations of matter.

This scenario remains an attractive possibility
for generating the baryon asymmetry in the universe.
We need now observables supporting
this scenario.
Three interesting questions come to
mind.

\noindent (1)\indent Is the excess of baryons 
produced in Leptogenesis consistent with the
observed amount
of matter relative to the photons observed in the
universe?\\
(2)\indent Are the neutrino oscillations and the
CP asymmetries observed at low energies related to
the CP in the early universe?\\
(3)\indent Are there remnants of this scenario at
cosmological scales that verify and support,
or perhaps contradict, Leptogenesis?

The answer to the first question is positive.  
Numerical studies have shown that a consistent
picture emerges provided that 
\begin{itemize}
\item[(i)] the dilution factor in the 
out--of--equilibrium decays is $D\sim10^{-3}$, and
\item[(ii)] the asymmetry $\varepsilon$ from 
individual decays is of order $10^{-4}$ to 
$10^{-5}$ and for $g_*=100$ degrees of freedom it gives the correct amount of matter relative to photons.
\end{itemize}
The answer to the second question is again positive.
There is now a flurry of activity with many models
proposed to relate the laboratory observations with
Leptogenesis.  I will discuss some of them in the
next section.

\section{Consequences of Leptogenesis}
Several groups developed models with massive neutrinos
which include neutrino oscillations, CP-violation
in the leptonic sector and the generation of a
lepton asymmetry.  A general approach includes the
standard model in a larger symmetry group and shows
that the parameters of the theory are consistent
with both the low energy phenomena and the
generation of a lepton--asymmetry 
\cite{ref11}--\cite{ref26}. Most of these 
models
incorporate the see--saw mechanism
\begin{equation}
m_{\nu}=m_D\, \frac{1}{M_R}\, m_D^T
\end{equation}
with $m_D$ the Dirac mass matrix and $M_R$
the mass matrix for the right--handed neutrinos.
We note that Dirac matrices that occur here and in
eq.~(6) are slightly different.  As we diagonalize
the right--handed mass matrix a unitary matrix
$U_R$ appears.  In the lepton--asymmetry occurs
the product
\begin{equation}
\tilde{m}_D = m_DU_R
\end{equation}
and thus the structure of the right--handed sector
influences the asymmetry.

Now many models imbed the standard model into a 
larger group and determine $m_D$ from the low energy
structure of the theory, including recent observations,
and deduce $U_R$ from the structure of the enlarged
theory.  The effort in this approach is to identify
general aspects common in several theories, like
SO(10), SU(5), Frogatt--Nielsen or texture models \cite{ref18,ref26}.
In several models the low energy phase from $m_D$ 
does not appear in the lepton asymmetry \cite{ref22}, 
which means
that the entire phase comes from the right--handed
sector.  However, there are models where the 
phases which are responsible for leptogenesis are
the ones that generate CP--violation at low energies.

A second approach enlarges the group to be 
left--right symmetric \cite{ref21}, i.e. $SU(2)_L\otimes SU_R(2)$.
In this case a Majorana mass term is also present
in the left--handed sector.  The mass relation is
now modified
\begin{equation}
m_{\nu}=m_ {LL}-m_D\, \frac{1}{M_R}\, m_D^T
\end{equation}
where $m_{LL}$ is a Majorana mass term for the
left--handed neutrinos.  The introduction of this
term is possible with the introduction of Higgs
triplets for the left-- and right--handed sectors
of the theory.  The mass matrices are given as
\begin{displaymath}
m_L=Fv_L\quad\quad{\rm and}\quad\quad m_R=Fv_R
\end{displaymath}
with $F$ the same matrix and $v_{L,R}$ the vacuum
expectation values of the triplet Higgses in the 
left and right sectors of the theory.  The fact 
that both mass matrices are proportional to the 
matrix $F$ simplifies the situation, because the
unitary matrices which diagonalize the mass matrices
for left--handed and right--handed fermions are now 
the same.

In case that the see--saw term in (14) is important
only the top quark contributes to the Dirac matrix
and a consistent solution was found.  Alternatively,
when $m_D$ is proportional to the charged
lepton mass matrix the see--saw term in eq.~(13) is
negligible, because $M_R$ is very heavy \cite{ref21}.
In this case the out--of--equilibrium condition is 
automatically fulfilled for typical values of the 
parameters.
In addition, the baryon asymmetry is of the correct 
magnitude and the large mixing angle solution for the 
solar problem is preferred.

In the laboratory, lepton number violation produces
neutrinoless double beta decay of certain nuclei
$(A,Z)\to(A,Z+2)+2e^-$ \cite{ref21,ref24}.  The decay width in these
extremely rare processes is proportional to the 
square of an effective neutrino mass 
\begin{displaymath}
\langle m_{ee}\rangle=\sum_i U^2_{ei}m_i
\end{displaymath}
originating from the Majorana sector.  Since the
matrix elements $U_{ei}$ are complex cancellations
can take place.  It is therefore interesting to ask
if the values of the parameters, which produce an
acceptable baryon asymmetry, also deliver a sizeable
$\langle m_{ee}\rangle$.  In the simple model, 
mentioned in the previous paragraph, the effective
neutrino mass $\langle m_{ee}\rangle$ can have 
values which are close to the experimental
bound on $\langle m_{ee}\rangle$ so that new 
experiments will be sensitive to neutrinoless
double--beta decay.

\section{CP Asymmetry in the Leptonic Sector}

In parallel to these activities there are serious
plans to measure CP violation in the leptonic sector.
The experiments will measure the difference
\begin{displaymath}
P(\nu_{\alpha}\to\nu_{\beta})-P(\bar{\nu}_{\alpha}\to
  \bar{\nu}_{\beta})
\end{displaymath}
which requires measuring $\nu_{\alpha}$, $\nu_{\beta}$,
and $\bar{\nu}_{\alpha}$, $\bar{\nu}_{\beta}$ 
interactions with hadrons at two different places.  This demands
precise knowledge of the neutrino--hadron cross sections
at low energies.  Even though neutrino--hadron 
interactions are more than 30 years old the low energy
cross sections are still poorly known.  In addition 
since the targets are very large, detailed instrumentation
of the detectors is not always possible.  For these
reasons there is an increasingly active community of
physicists concerned with these matters.
They are concerned in measuring the neutrino cross
sections carefully and at the same time developing 
accurate theoretical calculations.
The reactions of interest are 
quasi--elastic scattering, resonance production and 
the transition region
to deep inelastic scattering.  There is already an
effort in this direction with a conference organized
every year (NUINT 01 and 02).

A second aspect deals with the fact that the reactions 
occur in light and medium--heavy nuclei where additional
corrections are present.  Fig.~3 shows the 
neutrino nucleon cross section for $E_{\nu}<6$ GeV
from a Brookhaven experiment \cite{ref27}.  
The contributions of
quasi--elastic scattering and resonance production 
are clearly evident in the data up to $\sim 2$ GeV. They
produce the structure which looks like a step.  The
theoretical curve reproduces the data \cite{ref28}.  More work and
cross--checks will be necessary in order to obtain a
precise understanding of the data, required in order
to be able to observe CP--violation.  I presented only
one figure as an example, however, more results are
available and several studies are in progress 
\cite{ref29}.

%\Figure 3
\begin{figure}[htbp]
\epsfxsize=10cm
\centerline{\epsfbox{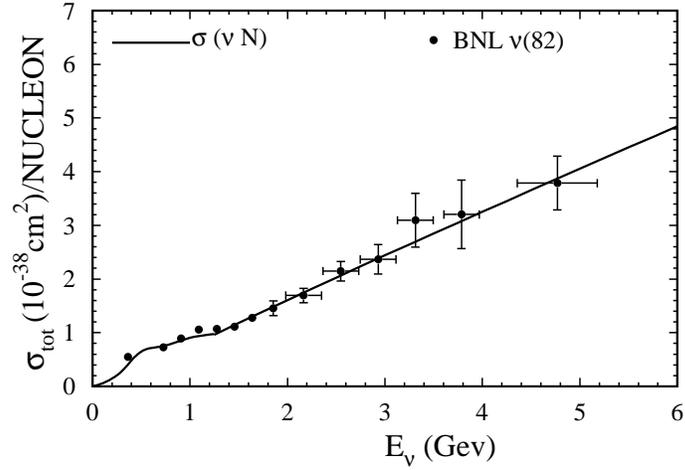}}
\caption{Neutrino nucleon cross section at low energies \protect\cite{ref27}.}
\label{fig:figure3}
\end{figure}
%\vspace{3.0cm}
%\begin{center} Fig.~3
%\end{center}

\section{Summary}
Leptogenesis presents a very attractive mechanism for
triggering the baryon asymmetry in the universe. It
has implications for the development of fluctuations
and inhomogeneities triggered by Majorana neutrinos
and later on by ordinary matter. This topic
is very interesting and deserves further investigation \cite{ref29}

A second development studies the connection between
the theoretical framework proposed for the heavy
Majorana states and laboratory phenomena including
neutrino oscillations.  Here there are many models
which try to find out common aspects and fulfill the
general conditions required by Leptogenesis.
A consistent picture has been developed in many models
and new developments are expected.

\section{Acknowledgment}
I wish to thank Dr. W. Rodejohann for discussions and Dr. A.
Srivastrava for correspondence concerning matter and lepton 
density fluctuations during the transition period.

\end{document}